\documentclass[a4paper]{llncs} 

\usepackage{acro}
\usepackage[utf8]{inputenc}
\usepackage[T1]{fontenc}
\usepackage{textcomp}
\usepackage{mathptmx}
\usepackage{fancyvrb}
\usepackage{algorithm}
\usepackage{algpseudocode}
\usepackage{mathtools}
\usepackage{bbm}
\usepackage[flushleft]{threeparttable}	
\usepackage{etoolbox}\AtBeginEnvironment{algorithmic}{\scriptsize} 
\usepackage[usenames,dvipsnames]{xcolor,colortbl}
\usepackage[colorlinks,allcolors=blue]{hyperref}
\usepackage{adjustbox}
\usepackage{amssymb}
\usepackage{pgfplots}
\usepackage{pgfplotstable}
\usepackage{filecontents}
\usepackage{dblfloatfix}    

\usepgfplotslibrary{colorbrewer}
\usetikzlibrary{pgfplots.statistics, pgfplots.colorbrewer} 

\pgfkeys{/pgf/number format/.cd,1000 sep={\,},fixed,precision=3}

\makeatletter
\pgfplotsset{
	boxplot/hide outliers/.code={
		\def\pgfplotsplothandlerboxplot@outlier{}%
	}
}
\makeatother

\newcolumntype{a}{>{\columncolor{Cyan!10!white}}c}
\newcolumntype{b}{>{\columncolor{Gray!10!white}}c}
\newcolumntype{w}{>{\columncolor{red!10!white}}c}
\newcolumntype{x}{>{\columncolor{green!10!white}}c}
\newcolumntype{y}{>{\columncolor{blue!10!white}}c}
\newcolumntype{z}{>{\columncolor{yellow!10!white}}c}

\usepgfplotslibrary{units}
\usetikzlibrary{arrows, calc, spy, patterns, backgrounds,decorations.pathreplacing, arrows.meta}

\pgfplotsset{
	my ybar legend/.style={
		legend image code/.code={
			\draw [##1] (0cm,-0.6ex) rectangle +(1.75em,1.1ex);
		},
	},
}

\makeatletter
\pgfdeclarepatternformonly[\hatchdistance,\hatchthickness]{flexible hatch}
{\pgfqpoint{0pt}{0pt}}
{\pgfqpoint{\hatchdistance}{\hatchdistance}}
{\pgfpoint{\hatchdistance-1pt}{\hatchdistance-1pt}}%
{
	\pgfsetcolor{\tikz@pattern@color}
	\pgfsetlinewidth{\hatchthickness}
	\pgfpathmoveto{\pgfqpoint{0pt}{0pt}}
	\pgfpathlineto{\pgfqpoint{\hatchdistance}{\hatchdistance}}
	\pgfusepath{stroke}
}

\makeatletter
\pgfplotsset{
	discontinuous line/.code={
		\pgfkeysalso{mesh, shorten <=#1, shorten >=#1,
			legend image code/.code={
				\draw [##1, shorten <=0cm] (0cm,0cm) -- (0.3cm,0cm);
				\draw [only marks] plot coordinates {(0.3cm,0cm)};
				\draw [##1, shorten >=0cm] (0.3cm,0cm) -- (0.6cm,0cm);
		}}
		\def\pgfplotsplothandlermesh@VISUALIZE@std@fill@andor@stroke{%
			\pgfplotspatchclass{\pgfplotsplothandlermesh@patchclass}{fill path}%
			\pgfplotsplothandlermesh@definecolor
			\pgfusepath{stroke}
			\pgfplotsplothandlermesh@show@normals@if@configured
		}%
	},
	discontinuous line/.default=1.5mm
}
\makeatother

\usepackage[firstinits=true,
uniquename=false,
uniquelist=false,
hyperref=auto,
maxbibnames=4,
maxcitenames=2,
style=numeric,
citestyle=numeric,
backref=false,
natbib=true,
style=trad-abbrv,
doi=true,
backend=biber]{biblatex}
\usepackage{booktabs}
\usepackage{sidecap}
\usepackage{graphicx, caption, subcaption}
\usepackage{relsize}
\usepackage[shrink=20]{microtype}
\usepackage{csquotes}
\usepackage{listings}
\usepackage{algpseudocode}
\usepackage[curves]{struktex}
\usepackage{mathrsfs}
\usepackage{multicol}

\usepackage{makeidx}  
\usepackage{icomma}
\usepackage{color}
\usepackage{array}
\usepackage{multirow}
\usepackage{layouts}
\usepackage{mathptmx}
\usepackage{breakurl}
\usepackage{comment}
\usepackage{ifthen}

\usepackage{amsmath}
\usepackage{amsfonts}
\usepackage{relsize}
\usepackage{xspace}

\usepackage{todonotes}
\usepackage{setspace}
\usepackage{cleveref}
\usepackage[binary-units=true]{siunitx}

\newcommand{\acrodef}[2]{\DeclareAcronym{#1}{short={#1},long={#2}}}
\acrodef{AABB}{axis-aligned bounding box}
\acrodef{API}{application programming interface}
\acrodef{ASIC}{application specific integrated circuit}
\acrodef{AST}{abstract syntax tree}
\acrodef{AVX}{advanced vector extensions}
\acrodef{BRAM}{block RAM}
\acrodef{CB}{compute-bound}
\acrodef{CER}{communication-to-execution ratio}
\acrodef{CG}{conjugate gradient}
\acrodef{ChebFD}{Chebyshev filter diagonalization}
\acrodef{CL}{cache line}
\acrodef{CoD}{cluster-on-die}
\acrodef{CPI}{cycles per instruction}
\acrodef{CPU}{central processing unit}
\acrodef{CUDA}{compute unified device architecture}
\acrodef{CST}{concrete syntax tree}
\acrodef{DP}{double precision}
\acrodef{DPM}{delay propagation mechanism}
\acrodef{DOF}{degree of freedoms}
\acrodef{DOM}{delay overlapping mechanism}
\acrodef{DPOM}{delay propagation and overlapping mechanisms}
\acrodef{DSL}{domain-specific language}
\acrodef{DVFS}{dynmic voltage frequency scaling}
\acrodef{ECM}{execution-cache-memory}
\acrodef{FD}{finite difference}
\acrodef{FEM}{finite element method}
\acrodef{FFC}{FEniCS Form Compiler}
\acrodef{FFT}{Fast Fourier transform}
\acrodef{FIFO}{first in first out}
\acrodef{FLOPS}{floating point operations per second}
\acrodef{FMA}{fused multiply-add}
\acrodef{FP}{floating-point}
\acrodef{FPGA}{field-programmable gate array}
\acrodef{FV}{finite volume}
\acrodef{GMRES}{generalized minimal residual}
\acrodef{GPU}{graphics processor unit}
\acrodef{GS}{Gauss-Seidel}
\acrodef{GUI}{graphical user interface}
\acrodef{HPCG}{High Performance Conjugate Gradient}
\acrodef{HDL}{hardware description language}
\acrodef{HHG}{hierarchical hybrid grid}
\acrodef{HLS}{high-level synthesis}
\acrodef{HPC}{high-performance computing}
\acrodef{IACA}{Intel Architecture Code Analyzer}
\acrodef{IP}{intellectual property}
\acrodef{ISA}{instruction set architecture}
\acrodef{ITAC}{Intel Trace Analyzer and Collector}
\acrodef{IR}{intermediate representation}
\acrodef{JIT}{just-in-time}
\acrodef{KPM}{Kernel Polynomial Method}
\acrodef{LC}{Layer Condition}
\acrodef{LFA}{local Fourier analysis}
\acrodef{LBM}{Lattice Boltzmann}
\acrodef{LLC}{last-level cache}
\acrodef{LoC}{lines of code}
\acrodef{LZR}{Leibniz Supercomputing Centre}
\acrodef{MB}{memory-bound}
\acrodef{MC}{memory controller}
\acrodef{MPI}{Message Passing Interface}
\acrodef{NDG}{nodal discontinuous Galerkin}
\acrodef{NDGTD}{nodal discontinuous Galerkin time domain}
\acrodef{NIC}{network interface controller}
\acrodef{OMP}{OpenMP}
\acrodef{NT}{non-temporal}
\acrodef{NUMA}{non-uniform memory access}
\acrodef{OS}{operating system}
\acrodef{OSACA}{Open-Source Architecture Code Analyzer}
\acrodef{P2P}{point-to-point}
\acrodef{PDE}{partial differential equation}
\acrodef{RAPL}{running average power limit}
\acrodef{PGAS}{partitioned global address space}
\acrodef{PPnR}{post place and route}
\acrodef{PPS}{processes per socket}
\acrodef{QDR}{quad data rate}
\acrodef{RAM}{random access memory}\acuse{RAM}
\acrodef{RBGS}{red-black Gauss-Seidel}
\acrodef{RDMA}{remote direct memory access}
\acrodef{RHS}{right-hand side}
\acrodef{RRZE}{Regional Computer Center Erlangen} 
\acrodef{RTL}{register transfer level}
\acrodef{SHM}{shared memory}
\acrodef{SPIR}{standard portable intermediate representation}
\acrodef{SPL}{software product lines}
\acrodef{SpMV}{sparse matrix-vector multiplication}
\acrodef{SIMD}{single instruction, multiple data}
\acrodef{SMP}{symmetric multiprocessing}
\acrodef{SMT}{simultaneous multithreading}
\acrodef{SP}{single precision}
\acrodef{SSE}{streaming SIMD extensions}
\acrodef{STL}{Standard Template Library}
\acrodef{TDP}{thermal design power}
\acrodef{TLB}{translation lookaside buffer}
\acrodef{TPDL}{target platform description language}
\acrodef{UFS}{Uncore frequency scaling}
\acrodef{WF}{wavefront}
\acrodef{XML}{eXtensible Markup Language}

\newcommand{\CPP}{C\nolinebreak[4]\hspace{-.05em}\raisebox{.23ex}{\relsize{-1}{++}}}

\newif\iftitle
\titletrue

\newcommand{\bq}{\begin{equation}}
\newcommand{\eq}{\end{equation}}
\newcommand{\bytes}{\mbox{B}}
\newcommand{\byte}{\mbox{byte}}
\newcommand{\second}{\mbox{s}}

\newcommand{\seconds}{\mbox{s}}

\newcommand{\bit}{\mbox{bit}}

\newcommand{\GHZ}{\mbox{GHz}}

\newcommand{\eos}{~.}
\newcommand{\cma}{~,}

\SaveVerb{Emmy}.Emmy.
\SaveVerb{Meggie}.Meggie.
\SaveVerb{HazelHen}.Hazel Hen.
\SaveVerb{SuperMUCNG}.SuperMUC-NG.
\SaveVerb{Hawk}.Hawk.
\SaveVerb{MPIcommsize}.MPI_Comm_size.
\SaveVerb{MPIwaitall}.MPI_Waitall.
\SaveVerb{MPIwait}.MPI_Wait.
\SaveVerb{MPIisend}.MPI_Isend.
\SaveVerb{MPIsend}.MPI_Send.
\SaveVerb{MPIirecv}.MPI_Irecv.
\SaveVerb{MPIrecv}.MPI_Recv.
\SaveVerb{MPIsendrecv}.MPI_Sendrecv.
\SaveVerb{MPIbarrier}.MPI_Barrier.
\SaveVerb{MPIallreduce}.MPI_Allreduce.
\SaveVerb{MPIreduce}.MPI_Reduce.
\SaveVerb{MPIalltoall}.MPI_Alltoall.
\SaveVerb{MPIallgather}.MPI_Allgather.
\SaveVerb{MPIgather}.MPI_Gather.
\SaveVerb{MPIscatter}.MPI_Scatter.
\SaveVerb{MPIbcast}.MPI_Bcast.
\SaveVerb{MPI}.MPI.
\SaveVerb{OpenMP}.OpenMP.
\SaveVerb{MPIOpenMP}.MPI+OpenMP.
\SaveVerb{NUMA}.NUMA.
\SaveVerb{STREAM}.STREAM.
\SaveVerb{Cosine}.Cosine.
\SaveVerb{DIV}.DIV.
\SaveVerb{vdivpd}.vdivpd.
\SaveVerb{SpMV}.SpMV.
\SaveVerb{static}.static.
\SaveVerb{LIKWID}.ToolA.
\SaveVerb{likwidperfctr}.featureA.
\SaveVerb{Chrono}.Chrono.
\SaveVerb{ccNUMA}.ccNUMA.
\SaveVerb{BLAS1}.BLAS-1.
\SaveVerb{BLAS2}.BLAS-2.

\definecolor{myblue}{RGB}{37,165,203}
\definecolor{myred}{RGB}{175,32,67}

\colorlet{ghcolor}{ProcessBlue}
\colorlet{aycolor}{RubineRed}
\newcommand{\AY}[2][]{\TODO{color=aycolor!30,#1}{AY}{#2}} 

\makeatletter
\renewcommand{\todo}[2][]{\@todo[caption={#2},#1]{\begin{spacing}{0.5}\fontfamily{phv}\fontseries{mc}\selectfont{#2\vspace{-1em}}\end{spacing}}}
\makeatother 
\newcommand{\TODO}[3]{\todo[#1]{\fontfamily{phv}\fontseries{mc}\fontsize{6.5pt}{6pt}\selectfont\textbf{\textcolor{black}{#2:\,}}\textcolor{black}{#3}}}

\newif\ifblind
\blindfalse
  
\begin{document}
\title{Analytic Modeling of Idle Waves in Parallel Programs: Communication, Cluster Topology, and Noise Impact}
	%

\ifblind
\author{Authors omitted for double-blind review process}
\else
\author{Ayesha Afzal\inst{1}, Georg Hager\inst{1}, and Gerhard Wellein\inst{1,2}}
\authorrunning{A.\ Afzal et al.} 
\institute{Erlangen Regional Computing Center (RRZE), 91058 Erlangen, Germany,\\
	\email{ayesha.afzal@fau.de, georg.hager@fau.de}
	\and
	Department of Computer Science, University of Erlangen-N\"urnberg, 91058 Erlangen, Germany,\\
	\email{gerhard.wellein@fau.de}}
\fi

\maketitle
\pgfkeys{/pgf/number format/.cd,1000 sep={\,}}

\begin{abstract}
Most distributed-memory bulk-synchronous parallel programs in HPC
assume that compute resources are available continuously and
homogeneously across the allocated set of compute nodes. However,
long one-off delays on individual processes can cause global
disturbances, so-called idle waves, by rippling through the system.
This process is mainly
governed by the communication topology of the underlying parallel
code. This paper makes significant contributions
to the understanding of idle wave
dynamics. We study the propagation mechanisms of idle
waves across the ranks of MPI-parallel programs. We present a
validated analytic model for their propagation velocity with respect
to communication parameters and topology, with a special emphasis on
sparse communication patterns. We study the interaction of idle
waves with MPI collectives and show that, depending on the
implementation, a collective may be transparent to the wave. Finally
we analyze two mechanisms of idle wave decay: topological decay,
which is rooted in differences in communication characteristics
among parts of the system, and noise-induced decay,
which is caused by system or application noise. We show that
noise-induced decay is largely independent of noise characteristics
but depends only on the overall noise power. An analytic expression
for idle wave decay rate with respect to noise power is derived.
For model validation we use microbenchmarks and stencil algorithms
on three different supercomputing platforms.
\end{abstract}

\section{Introduction}
\subsection{Idle waves in barrier-free bulk-synchronous
  parallel programs}

Parallel programs with alternating computation and communication
phases and without explicit synchronization are ubiquitous in high
performance computing. In theory, when running on a clean, undisturbed
system and lacking any load imbalance or other irregularities, such
applications should exhibit a regular lockstep pattern. In practice,
however, a variety of perturbations prevent this: system and network
noise, application imbalance, and delays caused by one-off events such
as administrative jobs, message re-transmits, I/O, etc.
Among all of these, long one-off events have the most immediate
impact on the regular compute-communicate pattern. They cause periods
of idleness in the process where they originated, but via inter-process
dependencies they ``ripple'' through the system and can thus
impact all other processes as well.
In massively parallel programs, delays can
occur anytime, impeding the performance of the application. On the other
hand, idle waves may also initiate desynchronization among
processes, which is not necessarily disadvantageous since it can
lead to automatic communication overlap~\cite{AfzalHW20}.

The speed and overall characteristics of idle wave propagation have
been the subject of some
scrutiny~\cite{markidis2015idle,peng2016idle,AfzalHW19,AfzalHW20}, but
a thorough analytical understanding of their dynamics with respect to
the communication topology of the underlying parallel code is still
lacking. There is also no investigation so far of the interaction of
idle waves with global operations such as reductions, and how the
system's hardware topology and the particular characteristics of
system noise impact the decay of idle waves. These topics will be
covered by the present work. We restrict ourselves to
\emph{process-scalable} scenarios, i.e., where multiple MPI
processes running on a hardware contention domain (such as
a memory interface or a shared out-level cache) do not feel
scalability loss due to hardware bottlenecks. 


\subsection{Related work} 

Noise has been studied for almost two decades. A large part of the work
focuses on sources of noise outside of the
control of the application and explores the influence of noise on collective
operations~\cite{nataraj2007ghost,ferreira2008characterizing,hoefler-loggopsim}.
However, it lacks coverage of pair-wise communication and
the interaction of noise with idle periods, which are 
common  in distributed-memory parallel codes. Gamell et
al.~\cite{Gamell:2015} noted the emergence of idle periods in the
context of failure recovery and failure masking of stencil codes.
Markidis et al.~\cite{markidis2015idle} used a LogGOPS
simulator~\cite{hoefler-loggopsim} to study idle waves
and postulated a linear wave equation
to describe wave propagation.

Afzal et al.~\cite{AfzalHW19,AfzalEuroMPI19Poster,AfzalHW20,AfzalHW21}
were the first to investigate the dynamics of idle waves,
(de)synchronization processes, and computational wavefront formation
in parallel programs with core-bound and memory-bound code, showing
that nonlinear processes dominate there.
Our work
builds on theirs to significantly extend it for analytic modeling with
further influence factors, such as communication
topology, communication concurrency, system topology and noise structure.

Significant prior work exists on the characterization of noise
and the influence of noise characteristics on performance of systems.
Ferreira et al.~\cite{ferreira2008characterizing} noted that HPC applications with
collectives can often absorb substantial amounts of high-frequency
noise, but tend to be affected by low-frequency noise.
Agarwal et al.~\cite{agarwal2005impact} found
noise properties to matter for the scalability of collectives,
comparing different distributions (exponential,
heavy tail, Bernoulli). 
Hoefler et al.~\cite{hoefler2010characterizing} used their
LogGOPS-based simulator and studied both \ac{P2P} and collective
operations. They found that application scalability is mostly
determined by the noise pattern and not the noise intensity.

In the context of idle wave propagation and decay, the present
work finds that the noise intensity is the main influence factor
rather that its detailed statistics.

\subsection{Contribution}
This work makes the following novel contributions:
\begin{itemize}
\item We analytically predict the propagation velocity of idle waves
  in scalable code with respect to (i) communication topology, i.e.,
  the distance and number of neighbors in point-to-point
  communication, and (ii) communication concurrency, i.e., how many
  point-to-point communications are grouped and subject to completion
  via \verb!MPI_Waitall!. 
\item The analytical model is validated with
  measurements on real systems and applied to microbenchmarks
  with synthetic communication topologies and a realistic scenario from
  the context of stencil codes with Cartesian domain decomposition.
\item We show that not all MPI collective routines eliminate a
  traveling idle wave; some may even be almost transparent
  to it, depending on their implementation. 
\item We show that idle wave decay can also be initiated by
  the system topology via inhomogeneities in point-to-point communication
  characteristics between MPI processes. 
\item We show analytically that the decay rate (and thus the survival
  time until running out) of an idle wave under the influence of noise
  is largely independent of the particular noise characteristics and
  depends only on the overall noise power. This prediction is
  validated with experiments.
\end{itemize}

\paragraph*{\textbf{Overview}}
This paper is organized as follows:
\Cref{sec:environment} provides details about
our experimental environment and methodology.
In~\Cref{sec:exec+comm}, we first introduce some
important terms to categorize 
execution and communication in distributed-memory parallel programs
and then develop and validate an analytical model 
of delay propagation.
\Cref{sec:collectives} covers the interaction of idle waves with collective primitives.
An analysis of idle wave decay  with respect to noise and system topology
is conducted in~\Cref{sec:decay}.
Finally, \Cref{sec:conclusion} concludes the paper and
gives an outlook to future work.	

\section{Test bed and experimental methods} \label{sec:environment}
The three clusters listed in Table~\ref{tab:system} were used to conduct various
experiments and validate our analytical models. 

\begin{table}[t]
	\caption{\small Key hardware and software specifications of systems.} 
	\label{tab:system}
	\begin{adjustbox}{width=0.985\textwidth}
		\begin{threeparttable}
			\setlength\extrarowheight{-0.7pt}
\setlength\tabcolsep{2pt}
\arrayrulecolor{Blue}
			\begin{tabular}[fragile]{c>{~}lwxy}
				\toprule
				\rowcolor[gray]{0.9}
				\cellcolor[gray]{0.9}&Systems  & \UseVerb{Emmy}\footnote{\ifblind{URL redacted for double-blind review}\else\url{https://anleitungen.rrze.fau.de/hpc/emmy-cluster}\fi}   &  \UseVerb{SuperMUCNG} & \UseVerb{Hawk}  \\
				\midrule
				\cellcolor[gray]{0.9}&Processor  & Intel Xeon Ivy Bridge EP   & Intel Xeon Skylake SP & AMD EPYC Rome    \\    
				\cellcolor[gray]{0.9}&Processor Model      & E5-2660 v2   & Platinum 8174 &  7742           \\
				\cellcolor[gray]{0.9}&Base clock speed &\SI{2.2}{\giga \Hz}    &  \SI{3.10}{\giga \Hz} (\SI{2.3}{\giga \Hz} used\mbox{$^\ast$})   & \SI{2.25}{\giga \Hz} \\
				\cellcolor[gray]{0.9}&Physical cores per node    & 20   & 48 & 128           \\
				\cellcolor[gray]{0.9}&Numa domains per node  &   2  & 2  &  8     \\
				\cellcolor[gray]{0.9}&LLC size & \SI{25}{\mega \byte}  & \SI{33}{\mega \byte}    &\SI{256}{\mega \byte} = 16 $\times$ \SI{16}{\mega \byte}  / CCX (4C)    \\
				\multirow{-7}{*}{\rotatebox{90}{\cellcolor[gray]{0.9} Micro-architecture}}&Memory per node (type)& \SI{64}{\giga \byte} (DDR3)  & \SI{96}{\giga \byte} (DDR4) & \SI{4}{\tera \byte} =16 $\times$ \SI{256}{\giga \byte} (DDR4)             \\
				
				\midrule
				\cellcolor[gray]{0.9}&Node interconnect    & QDR InfiniBand  & Omni-Path    & HDR InfiniBand         \\
				\cellcolor[gray]{0.9}&Interconnect topology & Fat-tree & Fat-tree & Enhanced 9D-Hypercube \\
				\multirow{-3}{*}{\rotatebox{90}{\cellcolor[gray]{0.9} Network}}&Raw bandwidth p. lnk n. dir &\SI{40}{\giga \bit \per \second}  &    \SI{100}{\giga \bit \per \second}  &    \SI{200}{\giga \bit \per \second}  \\
				
				\midrule
				\cellcolor[gray]{0.9}&Compiler    & Intel \CPP{} v2019.5.281      & Intel \CPP{} v2019.4.243   & Intel \CPP{} v2020.0.166     \\
				\cellcolor[gray]{0.9}&Optimization flags & -O3 -xHost  & -O3 -qopt-zmm-usage=high & -O3 -xHost \\
				\cellcolor[gray]{0.9}&SIMD & -xCORE-AVX2 &-xCORE-AVX512 & -mavx2\\
				\cellcolor[gray]{0.9}&Message passing library & Intel \verb.MPI. v2019u5     & Intel \verb.MPI. v2019u4    & Intel \verb.MPI. v2019u6       \\
				\multirow{-5}{*}{\rotatebox{90}{\cellcolor[gray]{0.9} Software}}&Operating system    & CentOS Linux v7.7.1908  &  SESU Linux ENT. Server 12 SP3    &  CentOS Linux 8.1.1911     \\
				
				\midrule
				\cellcolor[gray]{0.9}Tool&\verb.ITAC.    & v2019u4      & v2019  & v2020    \\
				\bottomrule
			\end{tabular}

			\begin{tablenotes}
				\small
				\item \mbox{$^\ast$} A power cap is applied on
				\UseVerb{SuperMUCNG}, i.e., the CPUs run by
				default on a lower than maximum clock speed (\SI{2.3}{\giga
					\Hz} instead of \SI{3.10}{\giga \Hz}).
			\end{tablenotes}
		\end{threeparttable}
	\end{adjustbox}
\end{table}
\footnotetext[3]{\ifblind{URL redacted for double-blind review}\else\url{https://anleitungen.rrze.fau.de/hpc/emmy-cluster}\fi}
Process-core affinity was enforced using the 
\texttt{I\_MPI\_PIN\_PROCESSOR\_LIST} environment variable.
We ignored the \ac{SMT} feature and used only physical cores. 
The clock frequency was always fixed to the base value of the respective CPUs
(or to 2.3\,\GHZ\ in case of \UseVerb{SuperMUCNG} because of
the power capping mechanism). 
On \UseVerb{Emmy}, experiments with up to 120 nodes were conducted 
on a set of nodes connected to seven 36-port leaf switches in order to
achieve homogeneous communication characteristics. A similar strategy
was not possible on the other systems.
Open-chain boundary conditions were employed unless specified otherwise. 
Communication delays for non-blocking calls were measured by time
spent in the \UseVerb{MPIwait} or \UseVerb{MPIwaitall} function.
We used
\ac{ITAC}\footnote{\url{https://software.intel.com/en-us/trace-analyzer}}
for timeline visualization and 
the \CPP{} high-resolution \UseVerb{Chrono} clock for timing measurements.
For tuning of the Intel MPI collectives implementations, we used the Intel
MPI \emph{autotuner}\footnote{\url{https://software.intel.com/content/www/us/en/develop/documentation/mpi-developer-reference-linux/top/environment-variable-reference/tuning-environment-variables/autotuning.html}
}; 
the configuration
space is defined by \verb.I_MPI_ADJUST_<opname>.\footnote{\url{https://software.intel.com/content/www/us/en/develop/documentation/mpi-developer-reference-windows/top/environment-variable-reference/i-mpi-adjust-family-environment-variables.html}
}.

We run barrier-free bulk-synchronous MPI-parallel micro-benchmarks 
with configurable latency-bound communication
and compute-bound workload. This results in process scalability,
i.e., there is no contention on memory interfaces, shared caches,
or network interfaces. 
The code loops over back-to-back divide instructions (\UseVerb{vdivpd}),
which have low but constant throughput.
The message size was set to \SI{1024}{\bytes}, which is well within the
default eager limit of the MPI implementation. For more realistic
workloads we chose a 3D Jacobi stencil and \ac{SpMV} with the \ac{HPCG}\footnote{\url{{https://www.hpcg-benchmark.org/}}} matrix.
Further characterization will be addressed in~\Cref{sec:exec+comm}.
One-off idle periods were generated by massively extending one computational
phase via doing extra work on one MPI rank, usually rank $5$.

All experiments described in this paper were conducted on all three benchmark
systems. However, we show the results for all of them only if there are
relevant differences.

\section{Idle wave propagation velocity for scalable code} \label{sec:exec+comm}

In this section we first categorize the execution and communication characteristics
of parallel applications. 
Later, we investigate how they influence the idle wave velocity and construct
an analytic model for the latter.

\subsection{Execution characteristics} \label{sec:exec}

HPC workloads have a wide spectrum of requirements regarding code execution
towards resources
of the parallel computing platform. The most straightforward categorization
is whether the workload is sensitive to certain resource bottlenecks,
such as memory bandwidth. Since we restrict ourselves
to scalable code here, we run the traditionally memory-bound algorithms such
as stencil updates or \ac{SpMV} with one MPI process per
contention domain (typically a ccNUMA node). This is not a problem
for the microbenchmarks since we deliberately choose an in-core
workload there.


\subsection{Categorization of communication characteristics} \label{sec:comm}

Here we briefly describe the different communication characteristics
under investigation. We start by assuming a \enquote{\ac{P2P}-homogeneous} situation where
all processes (except boundary processes in case of open boundary conditions)
have the same communication partners and characteristics.  We will later
lift this restriction and cover more general patterns.

\begin{figure}[t]
	\includegraphics{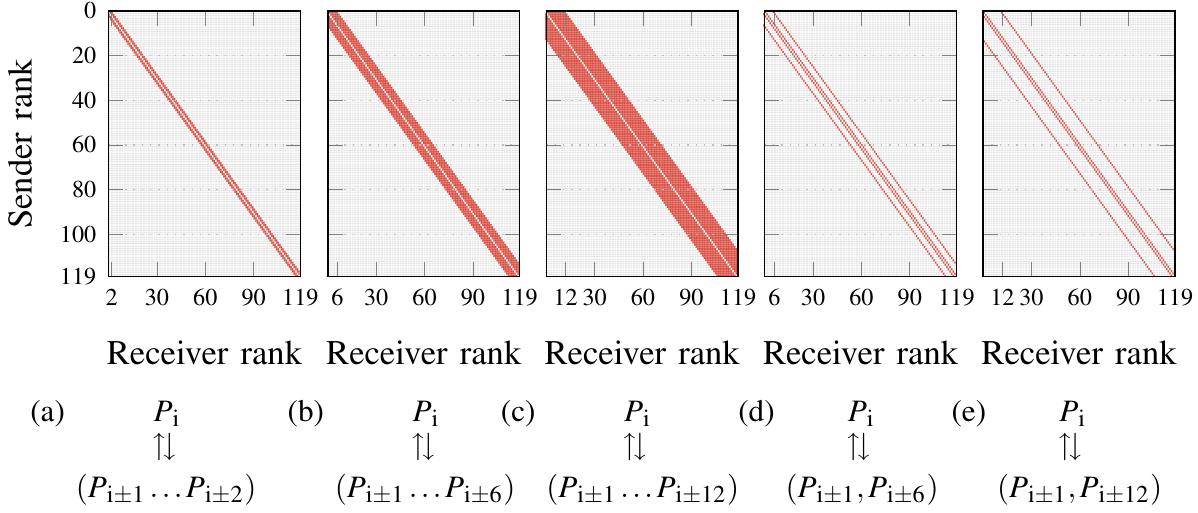}
	\caption{
		Compact and non-compact communication topologies with
		bidirectional open chain characteristics.
		$P_i$ sends (receives) 
		data to (from) $P_{i\pm1}$
		(a) till $P_{i\pm2}$ 
		(b) till $P_{i\pm6}$ 
		(c) till $P_{i\pm12}$, 
		(d) and $P_{i\pm6}$
		(e) and $P_{i\pm12}$.
	}
	\label{fig:Topology}
\end{figure}
\subsubsection{Communication topology}

Communication topology is a consequence of the physical problem underlying
the numerical method and of the algorithm (discretization, geometry). It boils down to the
question ``which other ranks does rank $i$ communicate with?'' and is
characterized by a \emph{topology matrix} (see~\Cref{fig:Topology} for examples
of \emph{compact} and \emph{noncompact} topologies).

In a compact topology, each process communicates with a dense,
continuous array of neighbors with distances $d = \pm 1, \pm 2,..., \pm j$.
The topology matrix comprises a dense band around the main diagonal.
In a noncompact topology, each process communicates with
processes that are not arranged as a continuous block, e.g.,
$d = \pm 1, \pm j$.
In both variants, the topology matrix can be symmetric or asymmetric.

For example, sparse matrices emerging from numerical algorithms with high locality
lead to compact communication structures, while stencil-like discretizations
on Cartesian grids lead to noncompact structures with far-outlying sub-diagonals.
Figures~\ref{fig:Topology}(a)--(c) depict symmetric cases with
$4$, $12$, and $24$ neighbors, respectively ($2$,
$6$ and $12$ distinct processes per direction) for every process,
while there are always four neighbors (two distinct processes per
direction) for both noncompact cases in Figures~\ref{fig:Topology}(d)--(e).
\begin{table}[t]
  \small\caption{Selected algorithms for communication concurrency in our MPI
    microbenchmarks. Arrows of the same color correspond to a single
    \protect\UseVerb{MPIwaitall} call. ``One distance'' means that
    one \protect\UseVerb{MPIwaitall} is responsible only for the send/recv pair
    of one particular communication distance, while ``all distances''
    means that it encompasses all distances in one dimension.
  }
  \label{tab:algorithm}
  \begin{adjustbox}{width=0.675\textwidth}
    \begin{threeparttable} 
      \includegraphics{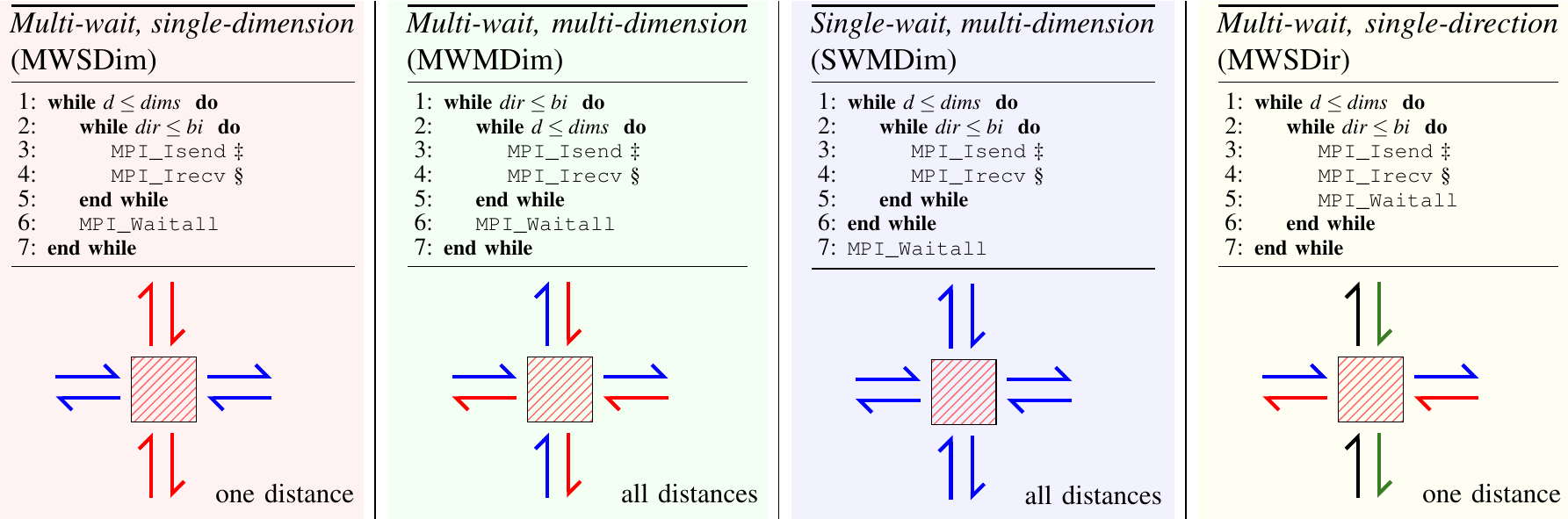}
    \end{threeparttable}
  \end{adjustbox}
  \begin{tablenotes}
    \scriptsize
  \item \mbox{$\ddagger$}  $P_i$ send to $P_{i+dir \times d}$; \hspace{0.2em}
    \mbox{$\S$} $P_i$ receive from $P_{i-dir \times d}$ 	
  \end{tablenotes}
\end{table}

\subsubsection{Communication concurrency} 

When a process communicates with others, it is often a deliberate choice
of the developer which communications are grouped together and later
finished using \UseVerb{MPIwaitall} (``split-waits'').
However, since interprocess dependencies have an impact on
idle wave propagation, such details are relevant. 
%
Of course, beyond user-defined
communication concurrency, there could still be nonconcurrency ``under
the hood,'' depending on the internals of the MPI implementation.

Here we restrict ourselves to a manageable subset of options that
nevertheless cover a substantial range of patterns.
We assume that all
\ac{P2P} communication is nonblocking.  Table~\ref{tab:algorithm}
shows the four variants covered here in a 2D Cartesian setting
according to the number of split-waits: 
\emph{multi-wait, single-dimension} (MWSDim),
\emph{multi-wait, multi-dimension} (MWMDim), 
\emph{single-wait, multi-dimension} (SWMDim), and \emph{multi-wait,
  single-direction} (MWSDir).
The iteration space of loops in~\Cref{tab:algorithm} is defined as the
outer (\verb.d.) loop goes over the Cartesian dimensions (i.e,
$x$ and $y$ here)
and the inner (\verb.dir.) loop goes over the two directions per
dimension (i.e., positive and negative).  For each direction
(e.g., positive $x$), the communication is effectively a linear
shift pattern; the pairing of send and receive operations per
\UseVerb{MPIwaitall} ensures that no deadlocks will occur.
The third and fourth option are corner cases with minimum
and maximum number of \UseVerb{MPIwaitall}s.

\subsubsection{More complex patterns}
Beyond the simple patterns described above, we will also cover more
general \emph{\ac{P2P} inhomogeneous} communication scenarios,
where subsets of processes have different communication
properties, such as in stencil codes or sparse-matrix algorithms.
\Cref{fig:Inhomo} shows an example with compact
long-range and short-range communication, which could emerge from
a sparse-matrix problem with ``fat'' and ``skinny'' regions of the matrix.
Finally, we will discuss  implementation alternatives of
collective communication primitives.


\subsection{Analytical model of idle wave propagation} \label{sec:model}
The propagation speed of an idle wave is the speed, in ranks per
second, with which it ripples through the system.  Previous studies of
idle wave mechanisms on silent systems~\cite{AfzalHW20,AfzalHW19}
characterized the influence of execution time, communication time,
communication characteristics (e.g., uni- vs.\ bidirectional
communication patterns and eager vs.\ rendezvous protocols), and the
number of active multi-threaded or single-threaded MPI processes on a
contended or noncontended domain.  However, the scope of that work was
restricted to a fixed \ac{P2P} communication pattern (fourth column in
\Cref{tab:algorithm} -- MWSDir).  Here we extend the analysis to more general
patterns, which show a much richer phenomenology. We restrict ourselves
to open boundary conditions across the MPI ranks. This is not
a severe limitation since it only affects the survival time and not
the propagation speed of the wave.

\paragraph{\textbf{Corner cases}}

Minimum idle wave speed (and thus maximum survival time) is
observed with simple direct next-neighbor
communication ($d=1$). If $T_\mathrm{exec}$ and
$T_\mathrm{comm}$ are execution and communication times of one iteration
of the bulk-synchronous program, then the idle wave speed is
\bq\label{eq:vmin}
  v^\mathrm{min}_\mathrm{silent} = 1 
  \left[\frac{\mbox{ranks}}{\mbox{iter}}\right]
  \times \frac{1}{T_\mathrm{exec}+T_\mathrm{comm}}
  \left[\frac{\mbox{iter}}{\mbox{s}}\right]\eos
\eq
In this case, the wave survives until it runs into system
boundaries~\cite{AfzalHW19}, i.e., for at most as many
time steps as there are MPI ranks. 
Barrier-like, i.e., long-distance synchronizing communication leads to
maximum speed and the wave dying out quickly in a minimum of one time
step.
Thus, in this case,
\bq\label{eq:vmax}
v^\mathrm{max}_\mathrm{silent} =
\alpha\left[\frac{\mbox{ranks}}{\mbox{iter}}\right]
\times \frac{1}{T_\mathrm{exec}+T_\mathrm{comm}}
  \left[\frac{\mbox{iter}}{\mbox{s}}\right]\cma
\eq
where $\alpha$ depends on the rank $r_\mathrm{inject}$ where the idle
wave originated:
\bq
\alpha = \max\left(\UseVerb{MPIcommsize}-
r_{\mathrm{inject}}-1, r_{\mathrm{inject}}-1\right)\eos
\eq

\paragraph{\textbf{Multi-neighbor communication}}

Away from the extreme cases, we have to distinguish between compact
and noncompact multi-neighbor communication patterns, but the
basic mechanisms are the same. The propagation speed of the idle
wave can be analytically modeled as
\bq\label{eq:propagationSpeed}
v_\mathrm{silent} = \kappa \cdot v^\mathrm{min}_\mathrm{silent}~
\left[\frac{\mbox{ranks}}{\mbox{s}}\right]\cma
\eq
Where $\kappa$ depends on communication concurrency and topology:
\bq\label{eq:kappa}
\kappa = \left\{
\begin{array}{cl}
  \displaystyle\sum_{k=1}^{j} k = \frac{j(j+1)}{2} & \mbox{~~if compact MWSDim / MWSDir / blocking}\\ 
  \displaystyle\sum_{k=1,j}^{} k = j+1 & \mbox{~~if non-compact MWSDim / MWSDir / blocking}\\
  j & \mbox{~~if MWMDim / SWMDim}
\end{array}
\right.\eos
\eq
Here, $j$ is the longest-distance communication partner of a rank.
Modifications to these expressions may apply for complex communication
topologies; we will discuss them in the validation section.


\begin{figure}[t]
  \includegraphics[scale=0.675]{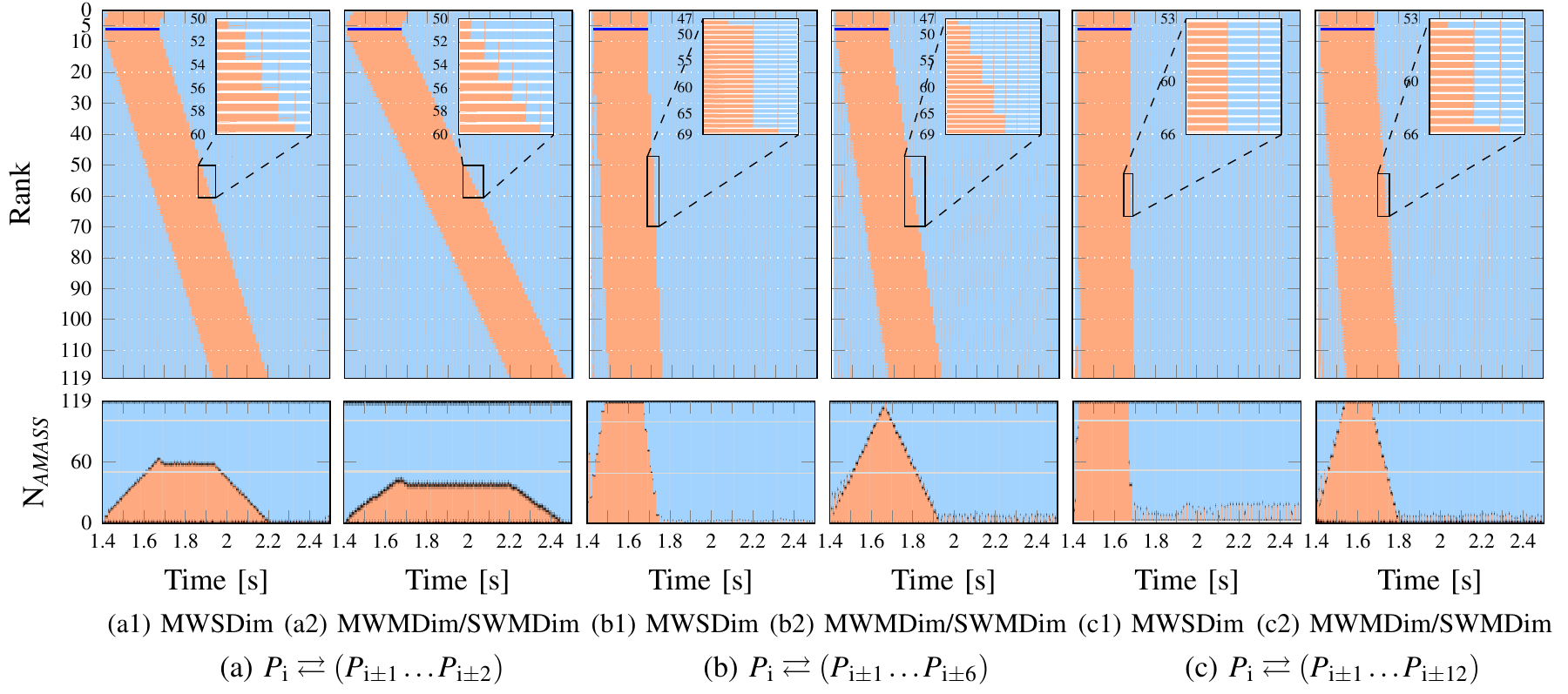}
  \caption{Top row: Idle wave propagation for 60 iterations
    in a core-bound microbenchmark
    for an injected delay at rank 5 (see text
    for details) and compact communication patterns with different numbers
    of communication partners: (a) two, (b) six, and (c) twelve
    partners per direction. The second row of panels shows the fraction
    of MPI ranks executing MPI library code.
}
  \label{fig:Compact} 
\end{figure}

\begin{figure}[t]
  \includegraphics[scale=0.675]{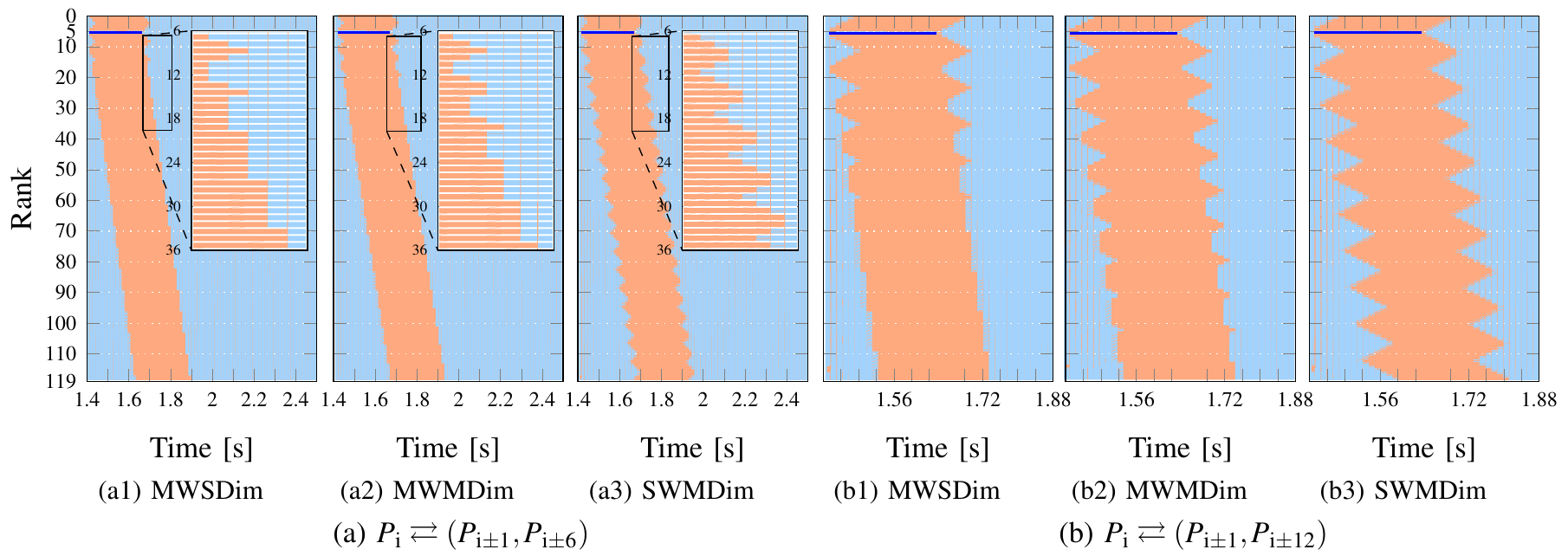}
  \caption{Idle wave propagation  in a core-bound microbenchmark
    for an injected delay at rank 5 (see text
    for details) and noncompact communication patterns with two
    communication partners per direction at different distances
    on \protect\UseVerb{Emmy}:
    (a) $P_i \rightleftarrows (P_{i\pm1}, P_{i\pm6})$ for 60
    iterations and (b) $P_i \rightleftarrows (P_{i\pm1}, P_{i\pm12})$ for
    20 iterations.
  }
  \label{fig:NonCompact}
\end{figure}

\subsection{Experimental validation} \label{sec:experiment}

In this section, we first validate the analytical model via measurements using
synthetic benchmarks on a real
system. Thereafter, we apply the model to a 3D a stencil code with
Cartesian domain decomposition.
Since stencil codes
are commonly memory-bound,  we run a single thread per
\verb.ccNUMA. domain only in order to maintain resource scalability.
Since the phenomenology matches across 
all three clusters (\Cref{tab:system}), we show results only
for the \UseVerb{Emmy} system.

\subsubsection{Microbenchmarks}

Figures~\ref{fig:Compact}~and~\ref{fig:NonCompact} (top row) show traces of the
propagation
of injected one-off idle phases (extra work at at rank 5, dark blue) and
its dependency on 
communication concurrency and communication topology, using the variants
shown in \Cref{tab:algorithm}.
In these experiments, we used an execution phase of
$T_\mathrm{exec}=\SI{13}{\milli \seconds}$ (light blue) and
a data volume of \SI{1}{\kibi\byte} per message.
The insets show close-ups of parts of the wave.
In the second row, a quantitative timeline of the number of MPI processes
executing MPI library code (i.e., waiting or communicating) is displayed.
In these settings, the natural system noise is weak enough to not cause
decay of the idle wave until it runs into the system boundary.

\paragraph{\textbf{Compact communication}}
In \Cref{fig:Compact}, the observed propagation speed of the
idle waves is independent of the number of split-waits, as expected.
Higher speeds are observed when (i) the overall communication distance
goes up, i.e., with growing number of communication partners, and
(ii) the number of dimensions spanned within each \UseVerb{MPIwaitall}
(communication concurrency).
In \Cref{fig:Compact}(a), where $P_i \rightleftarrows (P_{i\pm1},P_{i\pm2})$,
higher speed results in (a1) with $\kappa=\sum_{k=1}^{\mathrm{2}} k = 3$
due to the MWSDim concurrency pattern,
while in (a2) we have $\kappa= j = 2$ for the other patterns.
The data confirms the model
in (\ref{eq:propagationSpeed}) and (\ref{eq:kappa}).

In \Cref{fig:Compact}(b) and (c), the number of communication
partners per direction is increased to six and twelve, respectively,
with expected consequences: In (b1) we have
$\kappa=\sum_{k=1}^{\mathrm{6}} k = 21$, and in (b2) $\kappa= j =
6$. In (c1), we get$\kappa=\sum_{k=1}^{\mathrm{12}} k = 78$, confirming
intuitively our prediction that survival time in the high-speed limit
is equal to $T_\mathrm{exec}+T_\mathrm{comm}$. Finally, in
(c2) we get $\kappa= j = 12$.


The second row in \Cref{fig:Compact} shows that slower wave propagation
causes a more even spread of waiting times and thus resource
utilization across ranks.
A {ri\-sing}/{con\-stant}/falling slope indicates an oncoming/traveling/leaving wave.
Although our particular scenarios have been designed to show no
resource bottlenecks, these utilization shapes will be significant
in case of memory-bound execution or bandwidth-contended
communication~\cite{AfzalHW20}.
An exploration of these mechanisms is left for future work.

\paragraph{\textbf{Noncompact communication}}

Topology matrices with noncompact characteristics
(Figures~\ref{fig:Topology}(d)--(e)) entail a more complex
phenomenology of idle wave propagation. The presence of
``gaps'' leads to multiple waves propagating at different speeds,
with the added complication that each ``hop'' of a faster wave
sparks local idle waves wherever it hits (see \Cref{fig:NonCompact}).
These secondary
waves propagate and annihilate each other eventually (more specifically,
after $j/2$ hops), and
what remains is the fast wave emerging from the longest-distance
communication.
The speed of this residual wave is faster with (i) a larger number
of split-waits, (ii) a smaller number of communication dimensions
spanned by each \UseVerb{MPIwaitall}, and evidently (iii) a larger
longest communication distance $j$.

With respect to communication concurrency, there is a fundamental
difference between multiple split-waits and one
wait-for-all in non-compact communication.
The ``zig-zag'' pattern emerging from the two different propagation
speeds prevails in case of SWMDim (one
wait-for-all) but dies out for MWSDim and MWMDim 
after a couple of iterations.  This decay is entirely a consequence
of the communication concurrency and has nothing to do with
the other mechanisms of idle wave decay, such as
noise and communication inhomogeneity (see \Cref{sec:decay}).
The propagation of the ``envelope wave'' is untouched by
this effect.

This phenomenon is shown in \Cref{fig:NonCompact}(a1, b1, a2, b2), where
the zig-zag pattern dissolves eventually, and
the residual wave exhibits (a1) $\kappa=\sum_{k=1,6}^{} k = 7$,
(a2) $\kappa= j = 6$,
(b1) $\kappa=\sum_{k=1,6}^{} k = 13$,
and (b2) $\kappa= j = 12$.
The number of time
steps required for the zig-zag to even out depends
on the propagation speed.  In case of a single \UseVerb{MPIwaitall},
however (a3, b3), the pattern prevails. The envelope
travels with (a3) $\kappa= j = 6$ and
(b3) $\kappa= j = 12$. 

The results from these microbenchmarks show that our model is able
to describe the basic phenomenology of idle wave propagation
on a silent system in the parameter space under consideration. 
In the following we cover some more general patterns.

\begin{figure}[t]
	\begin{minipage}[c]{0.351\textwidth}
		\caption{Idle wave propagation with inhomogeneous compact communication
			charactersitics ($60$ iterations) on \protect\UseVerb{Emmy}. 
			(a) Topology matrix: $P_i$ sends (receives) \SI{1}{\kibi \byte} to
			(from) $P_{i\pm1}$,\ldots,$P_{i\pm3}$ for processes near boundaries
			and to (from) $P_{i\pm1}$,\ldots,$P_{i\pm12}$ for 40 inner processes.
			(b) Idle wave propagation for SWMDim concurrency.
		}
		\label{fig:Inhomo} 
	\end{minipage}
	\hspace{0.5em}
	\begin{minipage}[c]{0.639\textwidth}
		\resizebox{\columnwidth}{!}{%
			\includegraphics{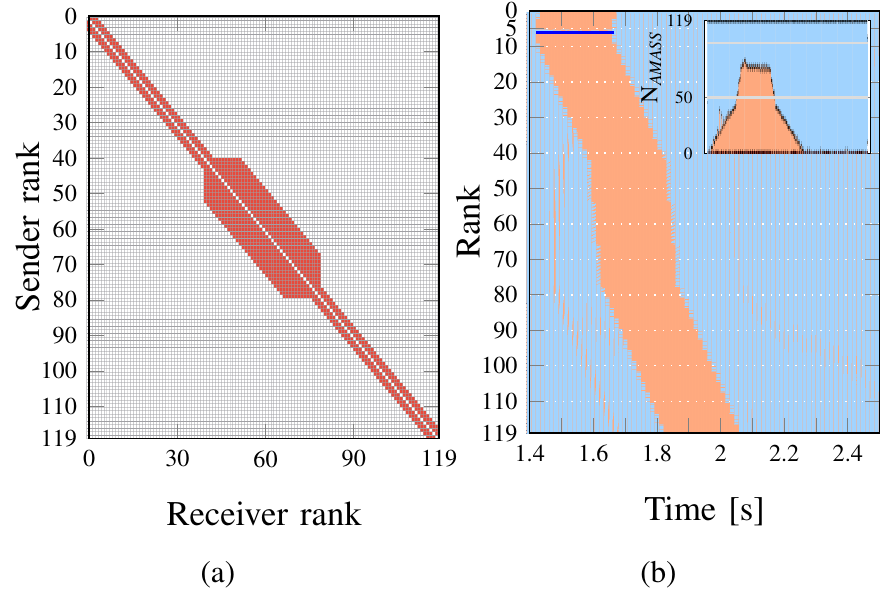}%
		}
	\end{minipage}
\end{figure}

\begin{figure}[t]
	\begin{minipage}[b]{0.46\textwidth} 
		\caption{Idle wave propagation within a double-precision
			3D Jacobi algorithm with Cartesian domain decomposition and
			bidirectional halo exchange ($15$ iterations) at a problem size of $1200^3$ and two
			different process grids ($120$ processes on \protect\UseVerb{Emmy})
			with open boundary conditions.
			Top row: topology matrices color-coded with communication volume.
			Bottom row: timelines of idle wave progression. 
			Orange color shows idleness in 
			\protect\UseVerb{MPIwait}, while pink color
			indicates waiting time in 
			\protect\UseVerb{MPIsend}. See text for communication
			grouping. 
			Single-message communication volumes are 
			(a) \SI{576}{\kilo \byte}, \SI{480}{\kilo \byte}, \SI{384}{\kilo \byte} and
			(b) \SI{960}{\kilo \byte}, \SI{576}{\kilo \byte}, \SI{192}{\kilo \byte} per
			dimension.}
		\label{fig:Jacobi} 
	\end{minipage}
	\hspace{0.9em}
	\begin{minipage}[b]{0.52\textwidth} 
		\resizebox{\columnwidth}{!}{%
			\includegraphics{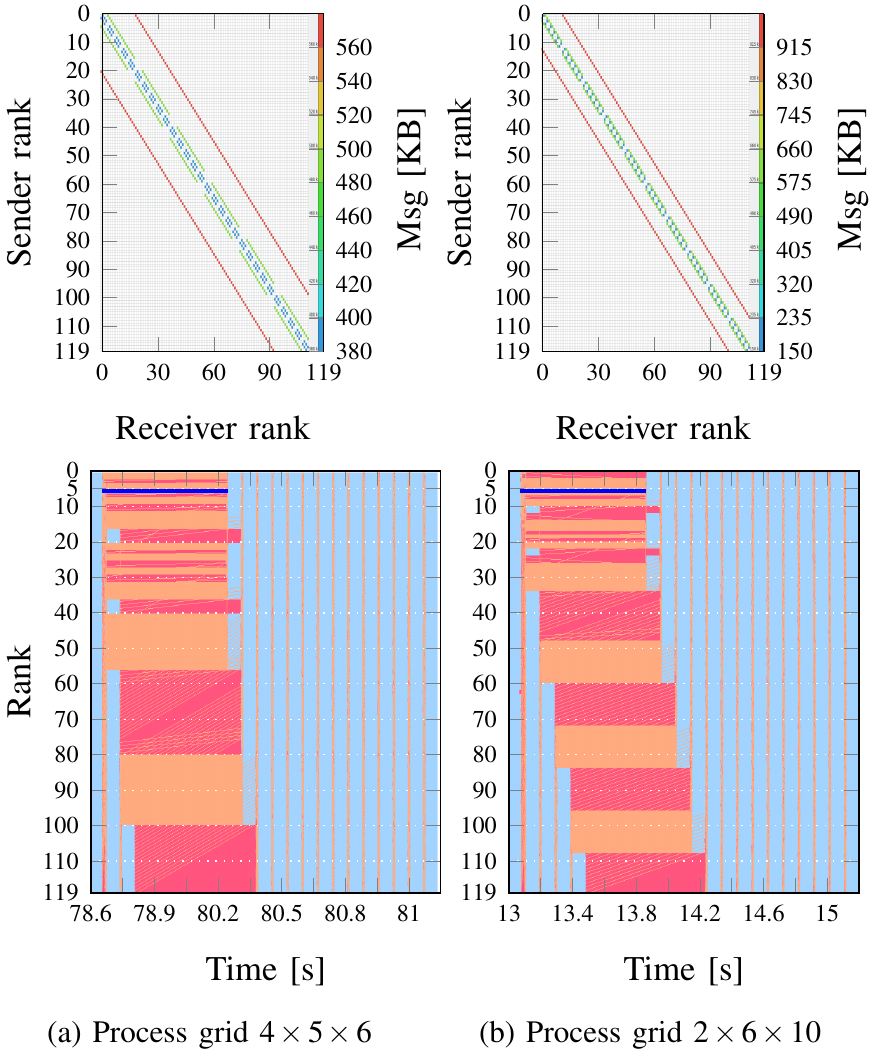}%
		}
	\end{minipage}
\end{figure}

\begin{figure}[t]
	\begin{minipage}[b]{0.36\textwidth}
		\caption{Idle wave propagation in sparse matrix-vector multiplication (\ac{SpMV})
			using the HPCG matrix with a problem size of $16^3$ per process 
			and bidirectional halo exchange ($15$ iterations) on \protect\UseVerb{Emmy}
			and three different process grids (a)--(c).
			Top row: topology matrices with color-coded communication volumes.
			Bottom row: Timelines of idle wave progression.
			Message sizes are \SI{8}{\bytes}, \SI{128}{\bytes},
			and \SI{2.05}{\kilo \bytes} per dimension (symmetry across main diagonals).}
		\label{fig:SpMVM}
	\end{minipage}
	\hspace{1.5em}
	\begin{minipage}[b]{0.55\textwidth}
		\resizebox{\columnwidth}{!}{%
			\includegraphics{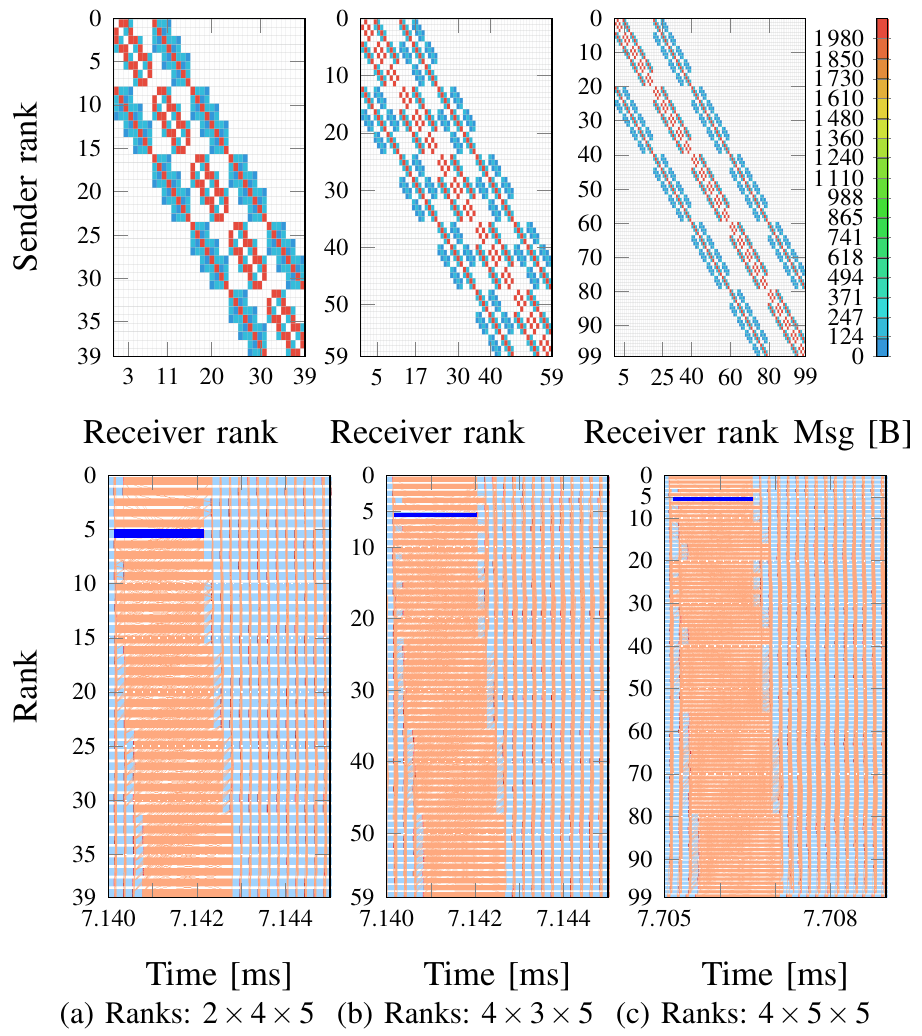}%
		}
	\end{minipage}
\end{figure}

\paragraph{\textbf{Inhomogeneous communication}}

From the basic propagation model and its validation on simple
communication patterns we can now advance to more complex scenarios.
In \Cref{fig:Inhomo}, we use a compact topology matrix that is
``fatter'' for the middle 40 processes, mimicking an inhomogeneous
situation that may, e.g., emerge with some sparse matrix problems (\Cref{fig:Inhomo}(a)).
Since the idle wave speed emerges from local properties of the
topology matrix, we expect a ``refraction effect,'' where
the wave travels faster within the fat region of the matrix.
Indeed, this is exactly what is observed (see \Cref{fig:Inhomo}(b)),
and the quantitative model of propagation speed holds for the
different regions: We have $\kappa= 12$ in the middle and $\kappa= 3$
elsewhere. 
%

\paragraph{\textbf{Blocking communication and eager vs. rendezvous mode}}
Instead of grouped nonblocking point-to-point calls, a popular choice is
\UseVerb{MPIsendrecv} for a pair of in- and outgoing messages along
the same direction. This is identical to the MWSDir case in \Cref{tab:algorithm},
so the phenomenology shown in Figures \ref{fig:Compact} (a1, b1, c1) and
Figures \ref{fig:NonCompact} (a1,b1) applies.
Similarly one can employ a \UseVerb{MPIirecv}/\UseVerb{MPIsend}/\UseVerb{MPIwait}
sequence within the innermost loop.
In all these cases, the wave propagation speed doubles in rendezvous
mode, where synchronization between sender and receiver is implied.
However, the difference between eager and rendezvous mode does not
impact the other variants beyond MWSDir.

\subsubsection{Stencil smoother with halo exchange}
\Cref{fig:Jacobi} shows an idle wave experiment with a
double-precision Jacobi smoother using Cartesian domain decomposition
and two different process grids (4$\times$5$\times$6
vs.\ 2$\times$6$\times$10; inner dimension goes first).
Here we used MWSDir concurrency via
\UseVerb{MPIirecv}/\UseVerb{MPIsend}/\UseVerb{MPIwait} per direction.
The message sizes are such that the rendezvous mode applies.
As expected from the model, the longest-distance communication
determines the overall wave speed, i.e., it is lower in case (b)
where the topology matrix is narrower. 

The communication topology is more intricate here than in the
microbenchmark studies covered so far. It turns out that all
connections apart from the longest-distance one can be summarized
by averaging over their respective distances and taking the largest
smaller integer (\verb.floor. function) when calculating
the $\kappa$ factor. For the case in \Cref{fig:Jacobi}(a), this
leads to $\kappa=2+20 = 22$, so the propagation speed is
$22\times 2 = 44$ times larger than $v^\mathrm{min}_\mathrm{silent}$.
For \Cref{fig:Jacobi}(b), we have $\kappa=0+12=12$ and thus
$24$ times $v^\mathrm{min}_\mathrm{silent}$. Both predictions are
confirmed by the data after the initial slow, short-distance
waves have died out. 



\subsubsection{SpMVM with halo exchange}
The High Performance Conjugate Gradient (HPCG) benchmark is popular
for ranking supercomputers beyond the ubiquitous LINPACK. 
Here we choose to discuss idle wave propagation during multiple
back-to-back sparse matrix-vector multiplications using the HPCG
matrix, which emerges from a sparse linear system using a $27$-point
stencil in 3D.
Communication is largely symmetric, except for boundaries.
The number of communication partners varies between $7$ (corners)
and $26$ (interior processes), and MWSDir concurrency applies
just like in the stencil example.
The per-process problem size is small enough for eager mode,
but communication time is a relevant contribution to the
overall runtime. 

\Cref{fig:SpMVM} shows idle wave propagation through three different
process grids with  $2 \times 4 \times 5=40$, $4 \times 3 \times 5=60$,
and $4 \times 5 \times 5=100$ ranks, respectively (inner dimension goes
first).
The decomposition is indicated in the captions of Figures~\ref{fig:SpMVM}(a)--(c).
In case (a) we get $\kappa = 8$, for (b) we get $\kappa = 12$, and
for (c) we get $\kappa = 24$.
\section{Idle waves interacting with MPI collectives}\label{sec:collectives}

Few MPI programs use point-to-point communications only. Concerning
idle wave propagation, the question arises which collective routines
may be transparent to a traveling wave. In practice, the elimination
or the survival of the wave may be desirable depending on the context;
for instance, it was shown that idle waves can lead to automatic
communication-computation overlap in desynchronized bottleneck-bound
programs~\cite{AfzalHW20}.

\begin{figure}[t]
  \includegraphics{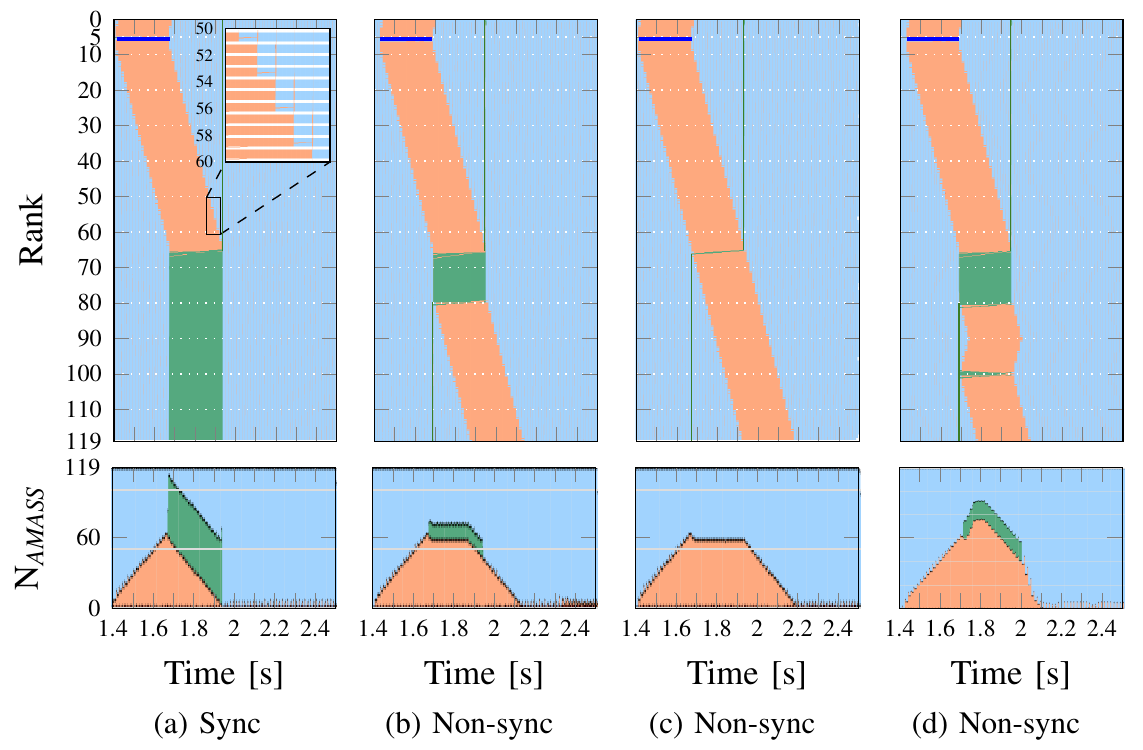}
  \caption{Transparency of collective routines for idle waves on \protect\UseVerb{Emmy}.
    (a) Default Intel MPI implementation of
    \protect\UseVerb{MPIallreduce} /
    \protect\UseVerb{MPIalltoall} /
    \protect\UseVerb{MPIallgather} /
    \protect\UseVerb{MPIscatter} /
    \protect\UseVerb{MPIbcast} /
    \protect\UseVerb{MPIbarrier} /
    \texttt{I\_MPI\_ADJUST\_REDUCE=1} /
    any collective with \texttt{I\_MPI\_TUNING\_AUTO\_SYNC=1},
    (b) default \protect\UseVerb{MPIreduce} or with
    \texttt{I\_MPI\_ADJUST\_REDUCE=8-11},
    (c) default \protect\UseVerb{MPIgather} / \protect\UseVerb{MPIreduce} with
    \texttt{I\_MPI\_ADJUST\_REDUCE=2,4-7},
    (d) \protect\UseVerb{MPIreduce} with \texttt{I\_MPI\_ADJUST\_REDUCE=3}.
    Collective calls are injected at rank 5 in the $20$th iteration
    and the root (where applicable) is rank $0$.
    The message size is \SI{1024}{\bytes}, and \texttt{MPI\_SUM}
    is used for all operations.
    Green color indicates the time spent by MPI processes in the collective routines.
  }
  \label{fig:Collectives}  
\end{figure}

The effects we discuss here are certainly heavily dependent on
the details of the MPI implementation, the communication buffer size,
and possibly other parameters, so it is impossible to give a comprehensive
overview. We thus restrict ourselves to Intel MPI on one of the three
benchmark systems (\UseVerb{Emmy}). The results are summarized in
\Cref{fig:Collectives} and discussed below.

\paragraph{\textbf{Globally synchronizing primitives}}

Examples of necessarily synchronizing collectives are
\UseVerb{MPIallreduce}, \UseVerb{MPIalltoall}, \UseVerb{MPIallgather},
\UseVerb{MPIbarrier}, etc.  These destroy propagating idle waves completely (see
~\Cref{fig:Collectives}(a)).
The default Intel implementations of
\protect\UseVerb{MPIscatter} and \protect\UseVerb{MPIbcast} are also
synchronizing.
If the \emph{autotuner mode} is enabled by setting \texttt{I\_MPI\_TUNING\_AUTO\_SYNC=1}
(disabled by default), an internal barrier
is called on every tuning iteration.
This, of course, completely eradicates an idle wave
on \emph{any} collective call.

\paragraph{\textbf{Global non-synchronizing primitives}}
\Cref{fig:Collectives}(b) shows an idle wave colliding with
the default Intel implementation of \UseVerb{MPIreduce}. Reductions
are not necessarily synchronizing, and indeed the idle wave
can pass the collective, which appears like a global, compact
communication block through which the wave travels with maximum
speed (see the discussion of inhomogeneous
communication above). 

If the survival of idle waves is desirable, one option is to avoid
synchronizing collectives if the performance implications are
noncritical.
In \Cref{fig:Collectives}(c), we show that the default 
\UseVerb{MPIgather} implementation is completely transparent
to the wave.

\paragraph{\textbf{Implementation variants}}
MPI implementations usually provide tuning knobs to optimize
the internal implementation of collectives in order to better
adapt it to the application. The process of finding the optimal
parameter settings can also be automated~\cite{vadhiyar2000automatically}.
With Intel MPI, the \verb.I_MPI_ADJUST_<opname>. environment
variable can be set to a value
that selects a particular implementation variant for the \verb.<opname>.
collective. Eleven documented settings are available in case of
\UseVerb{MPIreduce}.
\Cref{fig:Collectives}(c), although it depicts a
gather operation, is also applicable to \UseVerb{MPIreduce} with
\verb.I_MPI_ADJUST_REDUCE. set to $2$ or a value between $4$ and $7$.
Finally, \Cref{fig:Collectives}(d)
illustrates how the interaction of the idle wave with \UseVerb{MPIreduce}
changes for \verb.I_MPI_ADJUST_REDUCE. set to $3$ (topology-aware
Shumilin's algorithm).

Another option is to override the default shared-memory node-level
implementation of collectives and substitute it with a standard
point-to-point variant.
For instance, setting \verb.I_MPI_COLL_INTRANODE=pt2pt.
(insted of the default \verb.shm.) modifies the reduction
behavior from \Cref{fig:Collectives}(b) to \Cref{fig:Collectives}(c).

\section{Idle wave decay}\label{sec:decay}

The decay of traveling idle waves is a well-known
phenomenon~\cite{markidis2015idle}, and the underlying microscopic mechanism via
interaction with short idle periods (``noise'') is well understood~\cite{AfzalHW19}.
There are, however, two questions that have not been addressed so far:
(i) Does the system topology lead to idle wave decay also for
resource-scalable parallel programs?, and (ii) Which characteristics
of the system noise have an impact on the decay rate of the idle wave?
Here answer both.

\subsection{Topological decay}\label{sec:systemtopology}

\begin{figure}[!t]
	\includegraphics{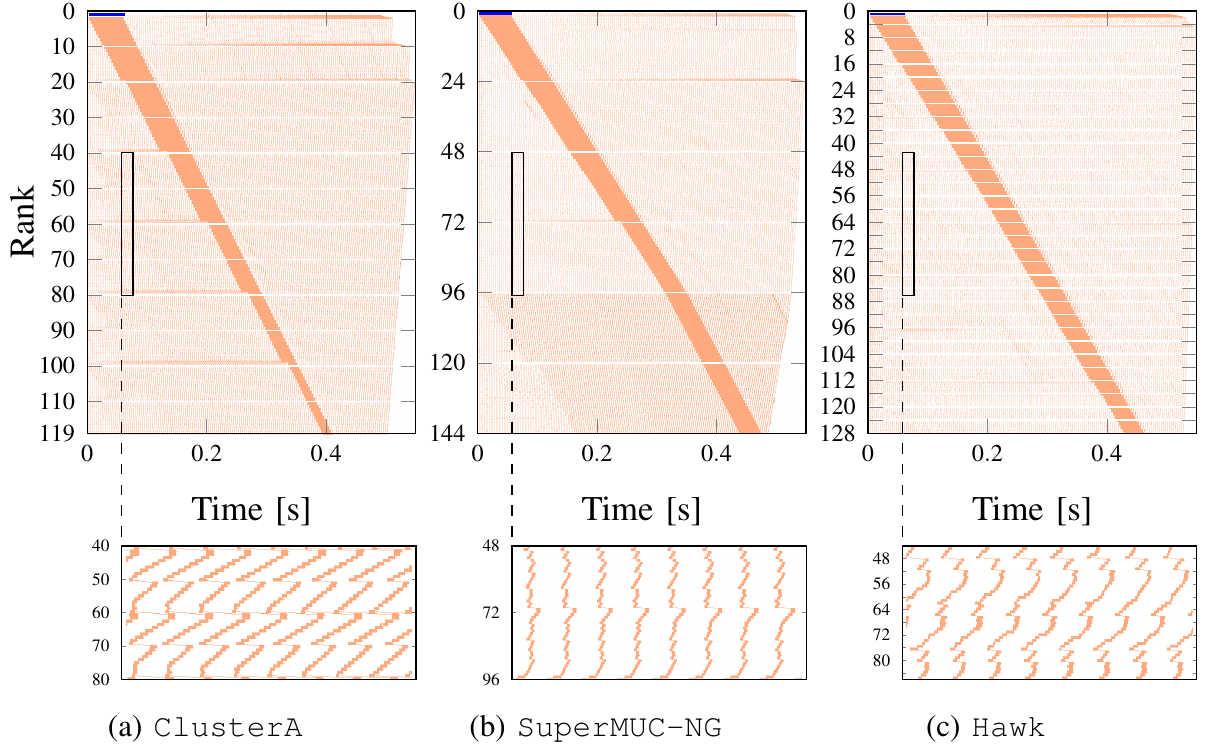}
	\caption{Topological idle wave decay on the benchmark systems
		running one process per core (scalable workload) using
		nonblocking MPI distance-1 communication topology
		(i.e., $P_i\rightleftarrows P_{i\pm 1}$) for $120$ iterations.
		We chose $T_\mathrm{exec}=\SI{2.7}{\milli \second}$ (white color) and
		injected extra work of \SI{58}{\milli \second} (blue color) at rank
		$0$. The message size was \SI{1}{\mega \bytes}.
		(a) $12$ domains (sockets), $120$ processes (b) $5$ domains
		(sockets), $120$ processes, (c) $30$ domains (CCX), $120$
		processes.  Topological boundaries exist at every $10$, $24$, and
		$4$ cores on \protect\UseVerb{Emmy}, \protect\UseVerb{SuperMUCNG} and
		\protect\UseVerb{Hawk}, respectively.}
	\label{fig:SystemTopology}
\end{figure}

It has been shown that the system topology, specifically a memory bandwidth
bottleneck, can cause idle wave decay without the presence
of system noise~\cite{AfzalHW20}. For the resource-scalable codes considered here
this mechanism does not apply, but there is more to system topology than
memory bottlenecks. The three benchmark systems we use here have quite
different features in this respect, even within a single node:
\UseVerb{Hawk} has $16$ cores ($4\times 4$ CCX) per ccNUMA domain,
  $4$ ccNUMA domains per socket, and $2$ sockets per node.
\UseVerb{SuperMUCNG} has $24$ cores  per ccNUMA domain, $1$ ccNUMA domain
  per socket, and $2$ sockets per node.
\UseVerb{Emmy} has $10$ cores per ccNUMA domain, $1$ ccNUMA domain
  per socket, and $2$ sockets per node.
The inherent topological boundaries cause communication inhomogeneities,
which create structured noise as small variations
in communication time (intranode vs.\ internode) propagate
and interact with the idle wave to
cause visible kinks. This is demonstrated in \Cref{fig:SystemTopology}
for the three benchmark clusters, running one MPI process per
ccNUMA domain. For 120 iterations, we measured an average
decay rate of \SI{149}{\micro \seconds \per rank} on
\UseVerb{SuperMUCNG}, \SI{203}{\micro \seconds \per rank} on
\UseVerb{Hawk}, and \SI{346}{\micro \seconds \per rank} on
\UseVerb{Emmy}. Although one might expect \UseVerb{Hawk} to show
the strongest topology effects due to its intricate node structure,
it is not only the number of hierarchy levels but also the actual
communication inhomogeneity that determines the decay effect.
In \Cref{fig:SystemTopology}, all 128 processes were run on a single
node of \UseVerb{Hawk}, so the internode boundary is missing there.

In order to substantiate the claim that this decay emerges from
system topology and communication inhomogeneities, we repeated
all experiments with \emph{round-robin placement} of MPI ranks across
nodes. In this way, node-level differences in communication
characteristics are all but eliminated since all interprocess boundaries
are internode boundaries. Indeed, the decay observed with standard
placement vanishes under these conditions.

\begin{figure}[t]
	\includegraphics[scale=0.675]{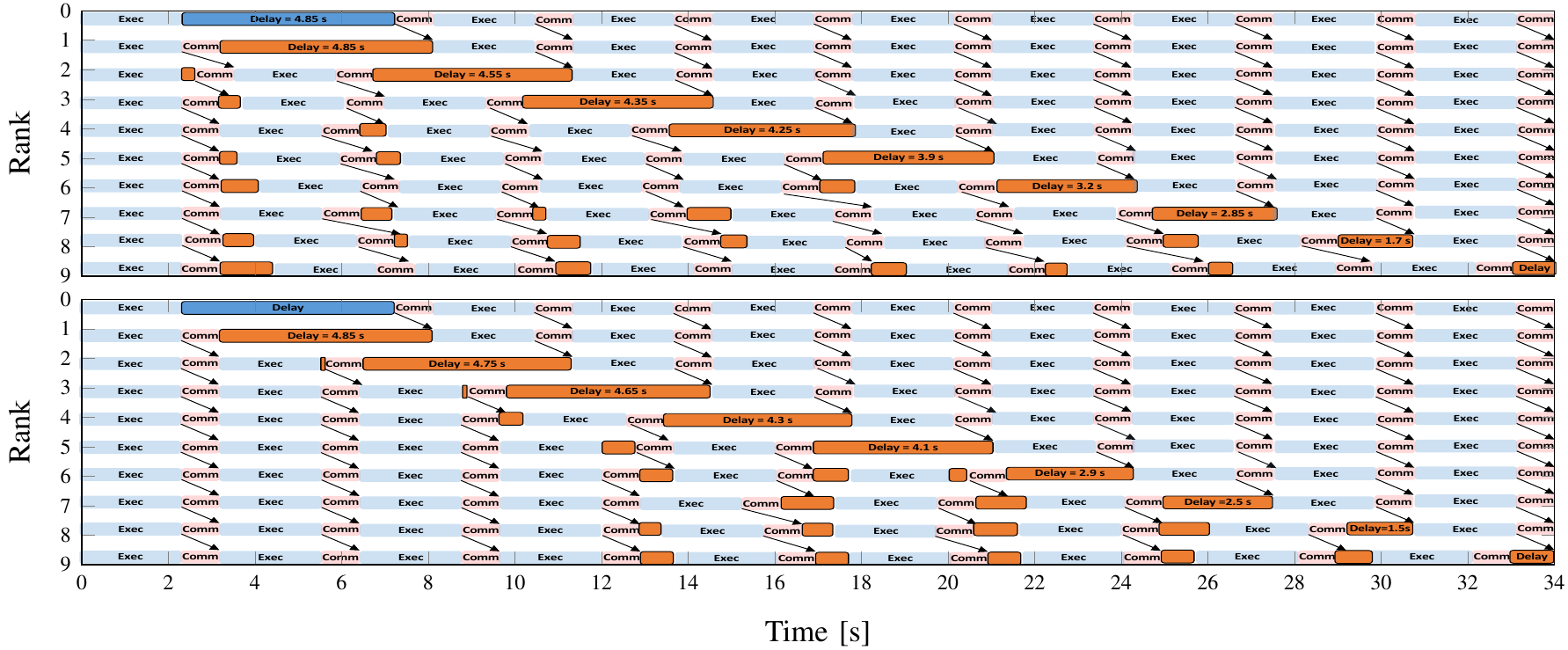}
	\caption{Experiment comparing the average decay rate
		of an idle wave (initial
		duration \SI{4850}{\milli\seconds}) for two different noise
		characteristics (top vs.\ bottom).
		In both cases, the integrated noise power is $9.1\%$ of the total area
		below the idle wave, i.e.,
		\SI{13}{\seconds} of \SI{142}{\seconds}, but the distribution
		of the fine-grained noise is different.
		However, the overall average decay rate is the same
		(\SI{480}{\milli\seconds\per rank}), as is the wave
		survival time (\SI{34}{\seconds}).
	}
	\label{fig:NoiseStatistics}
\end{figure}

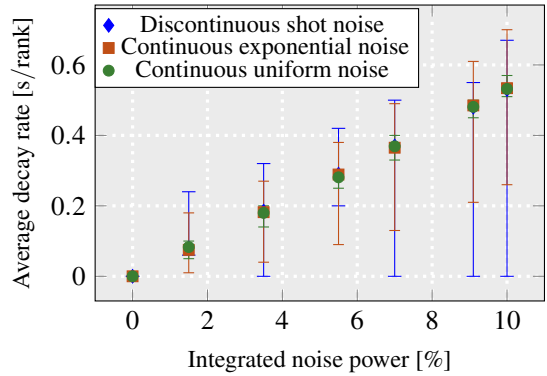
\begin{figure}[t]
	\begin{minipage}[c]{0.35\textwidth}
		\caption{Decay rate (min/max/median at
			sixteen cross-process transitions) of an idle period in
			\si{\seconds \per rank}, comparing three different noise patterns
			(see~\cite{AfzalHW19}) on
			the InfiniBand \protect\UseVerb{Emmy} (18 processes,
			one per node, single leaf switch).
			The $x$-axis shows
			integrated noise power with respect to overall integrated runtime
			of \SI{142}{\seconds}.
		}
		\label{fig:NoiseComparison} 
	\end{minipage}
	\hspace{0.8em}
	\begin{minipage}[c]{0.6\textwidth}
		\resizebox{\columnwidth}{!}{%
			\centering
		\begin{tikzpicture}
		\pgfplotstableread{figures/fig5-Noise/short.txt}\shotdata;		
		\pgfplotstableread{figures/fig5-Noise/exp.txt}\expdata;
		\pgfplotstableread{figures/fig5-Noise/uniform.txt}\uniformddata;
		\begin{axis}[trim axis left, trim axis right, scale only axis,
		axis background/.style={fill=white!92!black},
		width = 0.80\columnwidth,
		height = 0.2\textheight,
		xlabel = {Integrated noise power [\%]},
		ylabel = {Average decay rate [\si{\second \per rank}]},
		y label style={at={(0.05,0.5)}},
		x label style={font=\footnotesize},
		y label style={font=\footnotesize},
		legend style = {
			nodes={inner sep=0.04em}, 
			anchor=south, 
			font=\footnotesize,
			at={(0.36,0.72)}},
		ymajorgrids,
		y grid style={very thick, white, dotted},
		xmajorgrids,
		x grid style={very thick, white, dotted},
		legend columns = 1,
		]
		
		\addplot[only marks, mark=diamond*,mark size =2.5 pt,  blue, error bars/.cd, y dir=both, y explicit,]
		table
		[
		x expr=\thisrow{E}, 
		y error minus expr=\thisrow{Median}-\thisrow{Min},
		y error plus expr=\thisrow{Max}-\thisrow{Median}
		]{\shotdata};
		\addlegendentry{ Discontinuous shot noise}
		\addplot[only marks, mark=square*,mark size =2 pt, Bittersweet, error bars/.cd, y dir=both, y explicit,]
		table
		[
		x expr=\thisrow{E}, 
		y error minus expr=\thisrow{Median}-\thisrow{Min},
		y error plus expr=\thisrow{Max}-\thisrow{Median}
		]{\expdata};
		\addlegendentry{~Continuous exponential noise}
		\addplot[only marks, mark=*, OliveGreen,error bars/.cd, y dir=both, y explicit,]
		table
		[
		x expr=\thisrow{E}, 
		y error minus expr=\thisrow{Median}-\thisrow{Min},
		y error plus expr=\thisrow{Max}-\thisrow{Median}
		]{\uniformddata};
       	\addlegendentry{Continuous uniform noise}
		\end{axis}
		\end{tikzpicture}%
		}
	\end{minipage}
\end{figure}

\subsection{Noise-induced decay}\label{sec:noise}

For the purpose of this work, we define ``noise'' as any
(per-process) deviation from a fixed, repeatable, lockstep-type compute-communicate
pattern. In this sense, strong one-off delays are also noise,
but in this section we specifically consider noise that is
considerably more fine grained. 
One of the unsolved questions in previous work about idle wave
decay, specifically with resource-scalable code, is whether the
detailed statistical properties of the fine-grained noise
or just the integrated noise power impact the rate of decay. 
In order to exert full control over all noise characteristics,
we conduct experiments with artificial noise injections that are
orders of magnitude stronger than natural noise. Due to the
fundamental scale invariance of these mechanisms, the conclusions
must also hold for realistic scenarios.



How idle waves interact with each other in a nonlinear way has been
analyzed in previous work~\cite{AfzalHW19}; noise-induced decay is
just a variant of this process. Noise ``eats away'' at the trailing
edge of the wave, so a small idle period (i.e., a part of the noise) of
duration $T_\mathrm{noise}$ that collides with the idle wave shortens
the latter by an amount of exactly $T_\mathrm{noise}$.  This process
is cumulative, which leads to the immediate conclusion that multiple
interactions $\{T_\mathrm{noise}^i\}$ diminish the idle wave by
$\eta=\sum_iT_\mathrm{noise}^i$. Noise statistics is of minor importance
for the average decay rate. It will only impact the ``smoothness'' of
the decay.  \Cref{fig:NoiseStatistics} illustrates this
fact by comparing the decay of the same idle wave under two widely
different noise characteristics with identical integrated ``noise power''
$\eta$. Although the details of the decay are different, the
survival time and hence the average decay rate of the wave
is the same in both cases. This holds as long as 
the noise is fine-grained enough to not annihilate the idle wave in one
fell swoop at an early stage. 
Note that previous research~\cite{hoefler2010characterizing, agarwal2005impact,ferreira2008characterizing} only studied the influence of noise statistics
on application and global operations scalability.
Our observable is idle wave decay rate, which is largely robust
against noise statistics.



\subsubsection{Experimental validation}
To better validate this hypothesis, we measured the decay rate
of an idle wave under three  different noise characteristics with
the same noise power. \Cref{fig:NoiseComparison} shows results for
18 processes (one per node) on one leaf switch of \UseVerb{Emmy}
to rule out topological effects. Apart from this detail,
the setting is similar to \Cref{fig:NoiseStatistics}.
The microscopic shape of the decay is influenced by the statistics:
Shot noise,
i.e., random but strong, sparse noise injections of a single
duration, lead
to discontinuous decay and strong variations in decay rate
(diamonds in \Cref{fig:NoiseComparison}). On the other hand,
exponential (squares) and uniform (circles) noise characteristics,
where noise injections show a whole spectrum of durations, and
the variation in decay rates is much weaker.
The median of measured decay rates,
however, only depends on the noise power.  

%
%


\section{Summary and future work} \label{sec:conclusion}

We have presented an analytical model of idle wave propagation
speed based on communication topology and concurrency characteristics
of resource-scalable MPI programs. The model was validated
against simple microbenchmarks, a 3D stencil smoother, and sparse
matrix-vector multiplication with the HPCG matrix. We have also
shown that MPI collective routines can be transparent to
idle waves depending on the type and implementation of the
collective, which extends the relevance of idle wave phenomena
beyond bulk-synchronous algorithms without collective communication.
In light of the fact that the presence of idle waves is not
necessarily detrimental for performance, this result can be
quite relevant to the performance analysis of highly scalable
codes. Furthermore, we have uncovered the relevance of system topology
for idle wave decay: The presence of inhomogeneous communication
characteristics emerging from the hierarchical structure of modern
compute nodes leads to fine-grained noise that causes the decay
of idle waves. Finally, we have shown that it is the noise power,
and not its detailed statistical properties, that govern the
noise-induced decay rate. All these findings contribute significantly
to the understanding of the idle wave phenomenon on multicore clusters.

Future work will include the extension of the analysis to
programs that are not resource scalable, i.e., that are
limited by node-level or network-level bottlenecks. There
is also the open question which wave and noise phenomena
can be described by effective models that abstract away from
the details of the cluster hardware. Finally, we
will develop a capable MPI simulation tool that can take
node-level characteristics into account and will allow for
more extensive experimental studies and architectural exploration.

\ifblind
\else
\section*{Acknowledgments}
This work was supported by KONWIHR, the Bavarian Competence Network
for Scientific High Performance Computing in Bavaria, under project name ``OMI4papps,''
and by the BMBF under projects ``Metacca'' and ``SeASiTe.''
We are indebted to LRZ Garching and to HLRS Stuttgart for granting CPU hours on
their ``SuperMUC-NG'' and ``Hawk'' systems.  
\fi


\printbibliography

\end{document}